\documentclass{article}
\usepackage[T1]{fontenc} 
\usepackage[utf8]{inputenc} 
\usepackage{ismir,amsmath,cite,url}
\usepackage{graphicx}
\usepackage{color}

\usepackage{lineno}

\usepackage{caption}
\usepackage{listings}
\usepackage{placeins}
\usepackage{makecell}
\usepackage{multirow}

\usepackage{tikz}
\definecolor{pltorange}{rgb}{1,.498, .055}
\definecolor{pltblue}{rgb}{.122, .467, .706}

\title{DAC-JAX: A JAX Implementation of the Descript Audio Codec}

\oneauthor
 {David Braun} {Princeton University \\ Department of Computer Science \\ {\tt db1224@princeton.edu}}

\def\authorname{D. Braun}

\usepackage[bookmarks=false,pdfauthor={\authorname},pdfsubject={\papersubject},hidelinks]{hyperref}

\begin{document}

\maketitle

\begin{abstract}
We present an open-source implementation of the Descript Audio Codec (DAC) using Google's JAX ecosystem of Flax, Optax, Orbax, AUX, and CLU.
Our codebase enables the reuse of model weights from the original PyTorch DAC, and we confirm that the two implementations produce equivalent token sequences and decoded audio if given the same input.
We provide a training and fine-tuning script which supports device parallelism, although we have only verified it using brief training runs with a small dataset.
Even with limited GPU memory, the original DAC can compress or decompress a long audio file by processing it as a sequence of overlapping ``chunks.''
We implement this feature in JAX and benchmark the performance on two types of GPUs. On a consumer-grade GPU, DAC-JAX outperforms the original DAC for compression and decompression at all chunk sizes.
However, on a high-performance, cluster-based GPU,
DAC-JAX outperforms the original DAC for small chunk sizes but performs worse for large chunks.
\end{abstract}

\section{Introduction}

Descript Audio Codec (DAC), also known as Improved RVQGAN, is a high-fidelity neural audio codec~\cite{kumar2024high}. Similar codecs include Google's SoundStream~\cite{zeghidour2021soundstream} and Meta's EnCodec~\cite{defossez2022high}. These codecs all use residual vector quantization in an autoencoder, trained adversarially with a convolution-based discriminator.

Neural codecs like DAC have applications beyond traditional uses in digital telephony, streaming, and file compression. 
Researchers developed an automatic speech quality score and enhancement method based on a trained SoundStream model's quantization errors from encoding~\cite{fu2024self}.
The VALL-E text-to-speech system applies language modeling techniques to tokens from EnCodec~\cite{wang2023neural}. Language modeling has also been used to train music generation models, which are often conditioned on text.
For example, Google's \mbox{MusicLM}~\cite{agostinelli2023musiclm} builds on a variation of SoundStream, while Meta's MusicGen~\cite{copet2024simple} builds on a variation of AudioGen, and VampNet~\cite{garcia_vampnet_2023} builds on a variation of DAC. 

This paper introduces a DAC implementation in JAX. We verify that, given the same input, our implementation produces the same token sequences and reconstructed audio as the original PyTorch implementation.
Our project has three main goals.
First, it contributes to a discussion of the relative merits of the JAX and PyTorch ecosystems (\secref{implementation}). Second, we benchmark the PyTorch and JAX chunked DAC compression and decompression speeds in two GPU scenarios (\secref{benchmarks}).
Third, we discuss how downstream tasks (e.g., language model-based music generation) can benefit from JAX and its broader ecosystem (\secref{discussion}).
Our code is available under the MIT license.\footnote{\url{https://github.com/DBraun/DAC-JAX}}

\section{Implementation}\label{implementation}

This section begins by briefly describing the original DAC. Next it discusses the challenges and decision space, as well as some lessons learned, from implementing DAC in JAX.

\subsection{Properties of DAC}

DAC outperforms other codecs in subjective listening tests and objective metrics~\cite{kumar2024high}. 
It achieves equivalent quality audio reconstruction at a lower bitrate than previous codecs.
During compression, DAC's large model transforms 44.1 kHz audio into \textasciitilde86 (44100/512) Hz streams of tokens.\footnote{512 is the receptive field, which is the product of the ``encoder rates'' hyperparameters (2, 4, 8, 8). These parameters control the strides of 1D convolutional layers in the encoder. Note that the original DAC implementation describes this product as a ``hop size,'' not to be confused with the hop size of the chunked input.} 
The number of streams (codebooks) is 9; each codebook consists of 10-bit tokens; and each token has an 8-dimensional latent embedding. Multiplying (44100/512), 9, and 10 results in an effective bitrate per channel of~\textasciitilde7752 bps ($<$ 8 kbps).

One of DAC's advantages is its ability to compress long audio files and decompress the resulting token sequences using a constant GPU memory footprint. The key idea is to iterate over overlapping ``chunks'' of input audio.
Before the audio is split into chunks, it must be padded on both sides based on the aggregate delay introduced by the convolutional layers in DAC's encoder and decoder. The audio can then be split into overlapping chunks according to a window size and chunk hop size. 
A user-controllable parameter specifying the window size determines the GPU memory footprint.
This window size is converted to samples and rounded to an integer multiple of the receptive field.
The rounded window size, encoder rates, and decoder rates determine the chunk hop size used during compression. This hop size value appears later in the tables and charts in \secref{benchmarks}.
Larger window size implies larger hop size, which requires processing fewer chunks, and thus offers potentially higher throughput at the expense of higher latency.

\subsection{Flax}

Flax and Equinox are two of the most commonly used and actively maintained neural network architecture libraries that support JAX. We chose Flax over Equinox for three reasons. First, Flax adheres closely to the purely functional programming style of JAX: a model does not hold its own weights. In contrast, Equinox follows PyTorch's style in which a model holds weights. Second, although neither is an official Google product, Flax exists under Google's GitHub account, and Equinox does not. Finally, Faust-JAX (\secref{faust-jax}) already targets Flax. If DAC-JAX used Equinox instead, then Faust-JAX would need Equinox too, adding complexity. Although we picked Flax, Equinox has no deal-breakers, so we are not surprised to see SNAX use Equinox~\cite{SNAX}. SNAX is a JAX implementation of SNAC, which is a variation of DAC designed for compression at even lower bitrates\cite{Siuzdak_SNAC_Multi-Scale_Neural_2024}.

The functional design of Flax created a few challenges. First, the original DAC autoencoder dynamically changes the padding of its convolutional layers to zero during chunked compression and decompression. Flax modules cannot have side-effects like this. Our solution is to have the user instantiate DAC with a keyword argument of ``padding'' set to false if they want to use the model for chunked operations. Consequently, the instance of this model cannot be used for training. A second one, with ``padding'' by default set to true, would need to be instantiated, but the same weights could be used.

A second complication is that the PyTorch DAC autoencoder module can iterate over its submodules (e.g., convolutional layers, EncoderBlocks containing convolutional layers, and so on), but a Flax module cannot. This is important for programmatically determining the aggregate delay introduced by convolutional layers.
To overcome Flax's inability to iterate over submodules, we ended up writing static functions related to calculating delays and output lengths. Each custom layer of the architecture implements these static functions by explicitly composing calls to the same static functions of its submodules. Unfortunately, this results in code duplication: the same neural net architecture is composed in both these static functions and in the forward pass.

\begin{figure}[!b]
\begin{lstlisting}[language=Python, caption={PyTorch DAC's chunked compression and decompression.}, label=lst:pytorchcompression]
from dac.utils import load_model
model = load_model(model_type="44khz")
dac_file = model.compress(y)
y_hat = model.decompress(dac_file)
\end{lstlisting}
\end{figure}

The third design challenge involved implementing chunked compression and decompression.
To perform chunked compression and decompression over \texttt{y}---an instance of AudioSignal from AudioTools (\secref{audiotools})---one simply calls the code in Listing~\ref{lst:pytorchcompression}. In contrast, our JAX code requires a few more steps, as shown in Listing~\ref{lst:compression}.

\begin{figure*}[!ht]
\centering
\begin{lstlisting}[language=Python, caption={Chunked compression and decompression with JAX's JIT.}, label=lst:compression]
import jax
from dac_jax import load_model
model, variables = load_model(model_type="44khz", padding=False)

@jax.jit
def compress_chunk(x):
    return model.apply(variables, x, method="compress_chunk")

@jax.jit
def decompress_chunk(c):
    return model.apply(variables, c, method="decompress_chunk")

dac_file = model.compress(compress_chunk, y, 44_100)
y_hat = model.decompress(decompress_chunk, dac_file)
\end{lstlisting}
\end{figure*}

\subsection{ArgBind}

The original DAC used argbind\footnote{\url{https://github.com/pseeth/argbind/}} to configure training hyperparameters and set up command-line interfaces. Since argbind is neutral to PyTorch and JAX, we found it easy to continue using argbind with DAC-JAX. However, argbind has a few limitations, which are mentioned in its documentation. Considering these limitations, we believe that Meta's Hydra\footnote{\url{https://github.com/facebookresearch/hydra}} and Google's Gin Config\footnote{\url{https://github.com/google/gin-config}} are comparable alternatives to ArgBind.

\subsection{AudioTools}\label{audiotools}

Descript, the company behind DAC, also developed AudioTools, which is a library for object-oriented handling of audio signals.\footnote{\url{https://github.com/descriptinc/audiotools/}} An AudioSignal is an object with properties such as a sample rate, loudness, and spectrogram. Most properties only need to be computed if they are requested, and they are cached for later use. We began to implement AudioTools in JAX but abandoned it because we felt the object-oriented design didn't make sense for JAX's purely functional design. However, we speculate that it may make sense to adapt the AudioSignal class to a \texttt{flax.struct.dataclass} with fewer features.

AudioTools also has useful functions related to building datasets with data augmentations. Trying to avoid AudioTools's dependence on PyTorch, we instead implement our dataset loader with \texttt{tensorflow.data}\footnote{\url{https://tensorflow.org/guide/data}} and our augmentations with AUX (\secref{aux}).

\subsection{AUX}\label{aux}

Google developed PIX, an image-processing library in JAX that is useful for data augmentations. Similarly, Google developed AUX for audio-processing.\footnote{\url{https://github.com/google-deepmind/dm_aux/}} PIX is actively maintained, but AUX has not been updated since April 2022 (two years ago). Despite being concerned by this, we used AUX's discrete Fourier transform as a replacement for how AudioTools uses \texttt{torch.stft}. Originally, we wanted to use \texttt{jax.scipy.signal.stft} but encountered compilation warnings.\footnote{\url{https://github.com/google-deepmind/dm_aux/issues/2}} We hope that Google and outside researchers can give more attention to AUX in order to facilitate audio research in JAX.

\subsection{Julius and pyloudnorm}

For resampling audio from one sample rate to another, DAC uses Julius.\footnote{\url{https://github.com/adefossez/julius}} Since Julius is PyTorch-based, we implement the same resampling algorithm in JAX.

Normalizing the loudness of audio is an important part of pre-processing audio for training. To measure the loudness of audio signals, AudioTools implements pyloudnorm~\cite{steinmetz2021pyloudnorm} in PyTorch. For DAC-JAX, we use jaxloudnorm~\cite{jaxloudnorm}.

\subsection{Training and evaluation}

We provide a training script which supports device parallelism via JAX's \texttt{pmap} decorator. Furthermore, the data augmentations take place on each device rather than being CPU-bound by a primary dataloader. Our data augmentations include the volume normalization and data-balancing from the original DAC. However, we have not enabled the original DAC's random saliency augmentation, since we observed that it slowed down the dataloader.

We have not performed the same large training runs as DAC, but we have trained and fine-tuned on a small dataset. We implement metrics using the Google-developed Common Loop Utils (CLU) module. We use TensorBoard to inspect metrics and listen to reconstructed audio, and the results look correct.

The scripts directory of DAC has many functions related to evaluation, and we have not yet prioritized adapting them to JAX. Note that our training script re-implements the original DAC's loss functions and evaluation \textit{metrics}, but we didn't update all of the evaluation scripts.

\section{Benchmarks}\label{benchmarks}

Our codebase includes a script to measure DAC-JAX's chunked compression and decompression speeds. Our fork of the original DAC with an equivalent script is also online.\footnote{\url{https://github.com/DBraun/descript-audio-codec/tree/benchmarking}} All experiments use the 44.1 kHz 8 kbps model. We generate a one-item batch of short-duration audio at 44.1 kHz. We pass the audio to the compress function.
The compress function iterates over individual chunks, and we measure the time elapsed for the compression of each chunk.
The timing procedure is similar to decompression.
For 200 chunked executions, we measure the average and standard deviation of these execution times.

\begin{figure*}[t!]
  \centering
  \begin{minipage}{0.49\textwidth}
    \includegraphics[width=\linewidth]{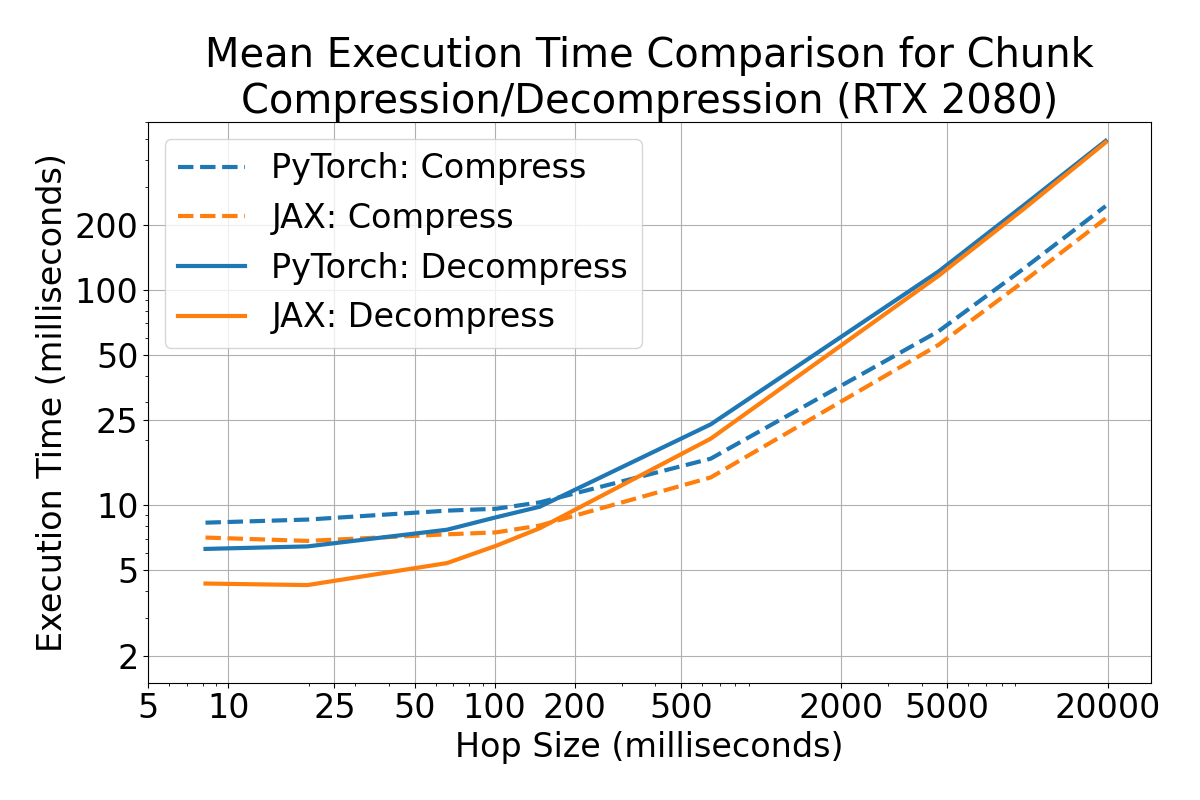}
    \vspace{-1.75\baselineskip}
    \caption{Log-log plot of the data from \tabref{tab:execution-2080}.}
    \label{fig:execution-2080}
  \end{minipage}
  \begin{minipage}{0.49\textwidth}
    \includegraphics[width=\linewidth]{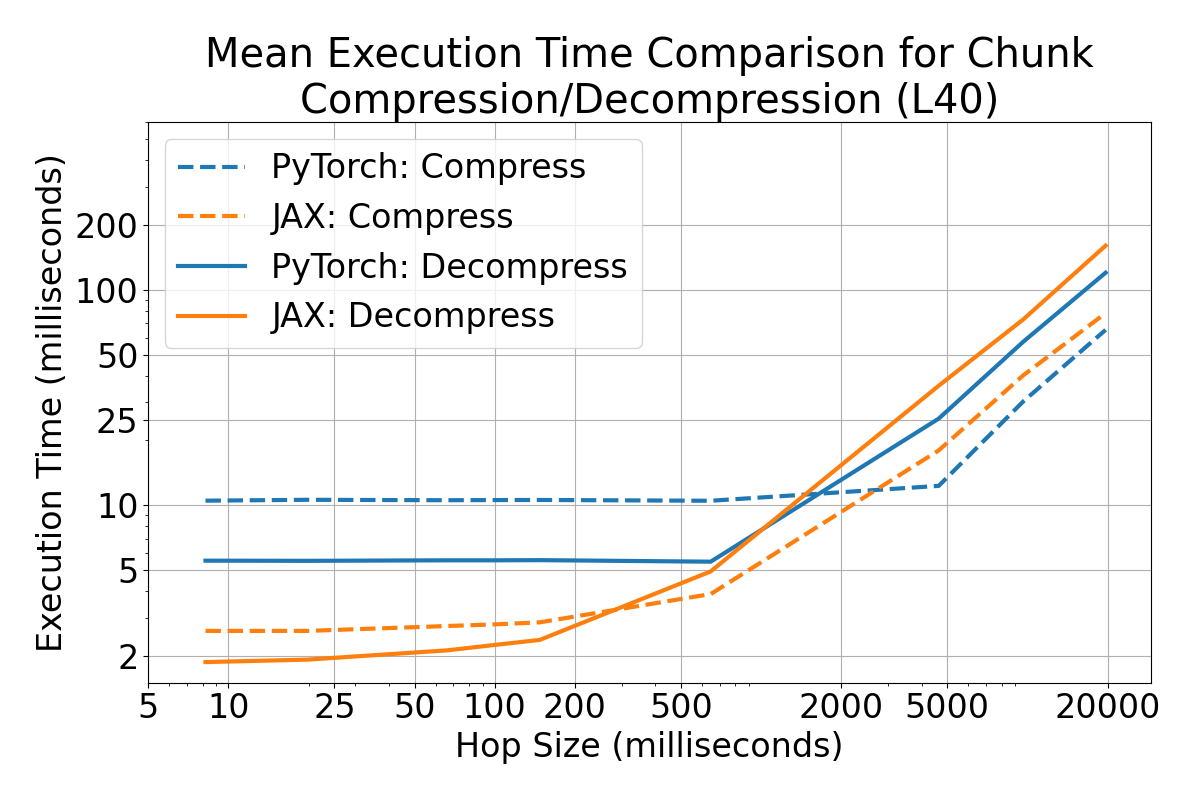}
    \vspace{-1.75\baselineskip}
    \caption{Log-log plot of the data from \tabref{tab:execution-l40}.}
    \label{fig:execution-l40}
  \end{minipage}\hfill 
  \vspace{0.5\baselineskip}
\end{figure*}

\figref{fig:execution-2080} shows these execution times for an Nvidia RTX 2080 GPU on a Windows 11 Pro workstation.
\figref{fig:execution-l40} shows the times for an Nvidia L40 GPU running on a remote Linux cluster.
In PyTorch, we use the official \texttt{torch.utils.benchmark.Timer} to measure the time of GPU operations.
To speed up the PyTorch DAC, we decorate its \texttt{compress} and \texttt{decompress} functions with \texttt{@torch.jit.export}, whereas our JAX functions have already been designed to use JIT.
For JAX, no official timer utility exists, so we use \texttt{timeit.repeat}. We call \texttt{block\_until\_ready} on the output to ensure that asynchronous GPU operations have completed.

\begin{table*}[ht!]
\small
\begin{center}
 \begin{tabular}{|r|r|r|r|r|r|r|}
  \hline
  \multirow{3}{*}{\makecell{Window Size\\ (ms)}} & \multirow{3}{*}{\makecell{Hop Size\\ (samples)}} & \multirow{3}{*}{\makecell{Hop Size\\ (ms)}} & \multicolumn{2}{c|}{PyTorch (ms)} & \multicolumn{2}{c|}{JAX (ms)} \\
  \cline{4-7}
  & & & \makecell{\centering \raisebox{-0.6ex}{Compress}} & \makecell{{\centering \raisebox{-0.6ex}{Decompress}}} & \makecell{{\centering \raisebox{-0.6ex}{Compress}}} & \makecell{{\centering \raisebox{-0.6ex}{Decompress}}} \\
  & & & \makecell{\centering \raisebox{0.4ex}{\tikz \draw[ultra thick, densely dashed, color=pltblue] (0,0) -- (1.5,0);}} & \makecell{\centering \raisebox{0.4ex}{\tikz \draw[ultra thick, color=pltblue] (0,0) -- (1.5,0);}} & \makecell{\centering \raisebox{0.4ex}{\tikz \draw[ultra thick, densely dashed, color=pltorange] (0,0) -- (1.5,0);}} & \makecell{\centering \raisebox{0.4ex}{\tikz \draw[ultra thick, color=pltorange] (0,0) -- (1.5,0);}} \\
  \hline
  370 & 362 & 8.2 & 8.30 (0.82) & 6.27 (0.32) & 7.08 (0.58) & 4.33 (2.75) \\
  \hline
  380 & 874 & 19.8 & 8.58 (1.24) & 6.44 (0.31) & 6.83 (0.59) & 4.26 (2.75) \\
  \hline
  420 & 2922 & 66.3 & 9.44 (0.88) & 7.70 (0.30) & 7.33 (0.53) & 5.39 (2.12) \\
  \hline
  460 & 4458 & 101.1 & 9.63 (0.85) & 8.79 (0.48) & 7.48 (0.65) & 6.48 (1.69) \\
  \hline
  500 & 6506 & 147.5 & 10.31 (1.31) & 9.84 (0.66) & 8.05 (0.66) & 7.82 (1.51) \\
  \hline
  1000 & 28522 & 646.8 & 16.45 (1.25) & 23.74 (1.25) & 13.46 (3.63) & 20.37 (2.22) \\
  \hline
  5000 & 204650 & 4640.6 & 64.30 (0.72) & 121.81 (1.33) & 55.61 (3.63) & 116.32 (14.72) \\
  \hline
  10000 & 425322 & 9644.5 & 125.07 (1.54) & 245.90 (2.48) & 109.08 (8.93) & 236.57 (31.26) \\
  \hline
  20000 & 866154 & 19640.7 & 244.64 (0.80) & 491.34 (1.00) & 214.42 (14.66) & 483.58 (64.58) \\
  \hline
 \end{tabular}
 \caption{Mean compression and decompression times (standard deviation in parentheses) on a Windows 11 Pro PC with an Nvidia RTX 2080 GPU, for varying hop sizes. 
 }
 \label{tab:execution-2080}
\end{center}
\end{table*}

\begin{table*}[ht!]
\small
\begin{center}
 \begin{tabular}{|r|r|r|r|r|r|r|}
  \hline
  \multirow{3}{*}{\makecell{Window Size\\ (ms)}} & \multirow{3}{*}{\makecell{Hop Size\\ (samples)}} & \multirow{3}{*}{\makecell{Hop Size\\ (ms)}} & \multicolumn{2}{c|}{PyTorch (ms)} & \multicolumn{2}{c|}{JAX (ms)} \\
  \cline{4-7}
  & & & \makecell{\centering \raisebox{-0.6ex}{Compress}} & \makecell{{\centering \raisebox{-0.6ex}{Decompress}}} & \makecell{{\centering \raisebox{-0.6ex}{Compress}}} & \makecell{{\centering \raisebox{-0.6ex}{Decompress}}} \\
  & & & \makecell{\centering \raisebox{0.4ex}{\tikz \draw[ultra thick, densely dashed, color=pltblue] (0,0) -- (1.5,0);}} & \makecell{\centering \raisebox{0.4ex}{\tikz \draw[ultra thick, color=pltblue] (0,0) -- (1.5,0);}} & \makecell{\centering \raisebox{0.4ex}{\tikz \draw[ultra thick, densely dashed, color=pltorange] (0,0) -- (1.5,0);}} & \makecell{\centering \raisebox{0.4ex}{\tikz \draw[ultra thick, color=pltorange] (0,0) -- (1.5,0);}} \\
  \hline
    370 & 362 & 8.2 & 10.50 (0.32) & 5.53 (0.15) & 2.61 (0.04) & 1.87 (0.83) \\
    \hline
    380 & 874 & 19.8 & 10.60 (0.30) & 5.52 (0.14) & 2.61 (0.04) & 1.92 (0.88) \\
    \hline
    420 & 2922 & 66.3 & 10.55 (0.28) & 5.55 (0.15) & 2.75 (0.05) & 2.12 (0.52) \\
    \hline
    460 & 4458 & 101.1 & 10.57 (0.29) & 5.55 (0.16) & 2.80 (0.05) & 2.25 (0.39) \\
    \hline
    500 & 6506 & 147.5 & 10.58 (0.30) & 5.56 (0.15) & 2.86 (0.06) & 2.37 (0.29) \\
    \hline
    1000 & 28522 & 646.8 & 10.49 (0.28) & 5.47 (0.15) & 3.87 (0.11) & 4.93 (0.24) \\
    \hline
    5000 & 204650 & 4640.6 & 12.30 (0.07) & 25.24 (0.14) & 17.97 (1.00) & 35.87 (3.00) \\
    \hline
    10000 & 425322 & 9644.5 & 30.35 (0.11) & 57.46 (0.21) & 40.11 (1.86) & 72.93 (7.57) \\
    \hline
    20000 & 866154 & 19640.7 & 65.19 (0.13) & 120.08 (0.94) & 77.99 (4.52) & 160.36 (18.09) \\
  \hline
 \end{tabular}
 \caption{Mean compression and decompression times on a Linux cluster with an Nvidia L40 GPU, for varying hop sizes.
 }
 \label{tab:execution-l40}
\end{center}
\end{table*}

\begin{figure*}[ht!]
  \centering
  \begin{minipage}{0.49\textwidth}
    \includegraphics[width=\linewidth]{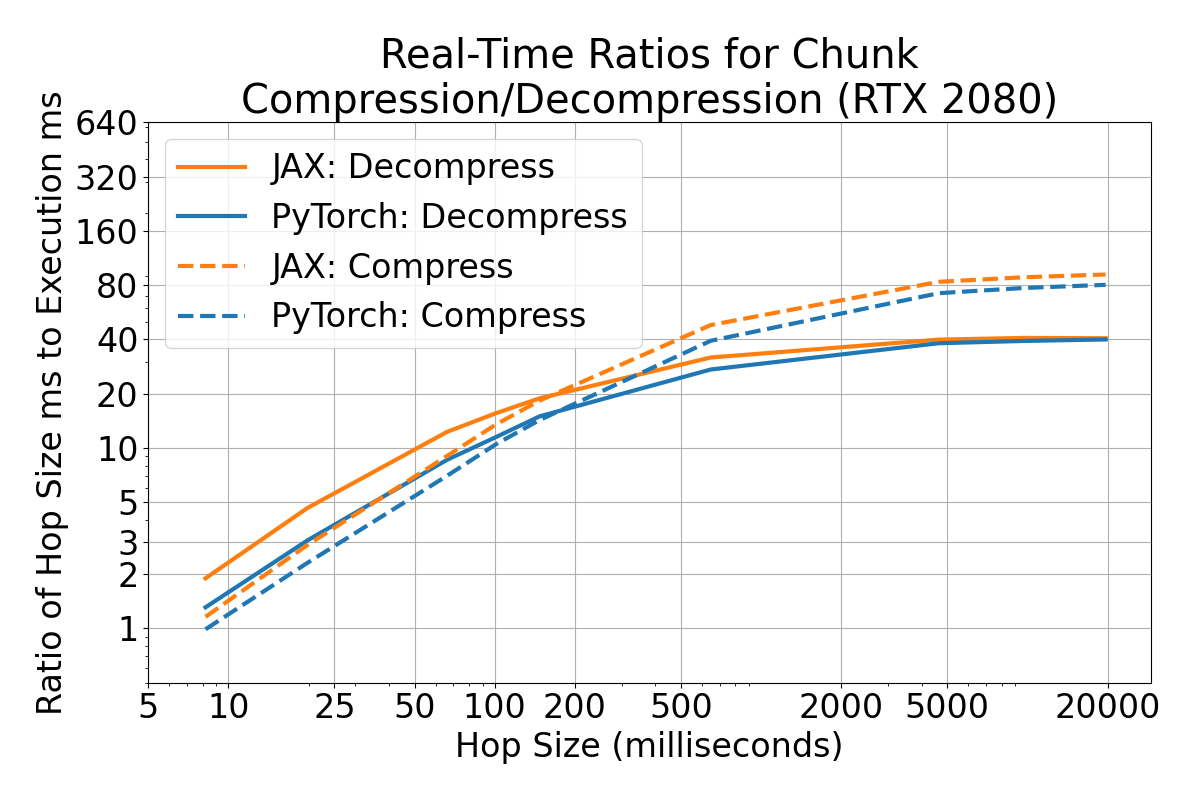}
  \end{minipage}\hfill 
  \begin{minipage}{0.49\textwidth}
    \includegraphics[width=\linewidth]{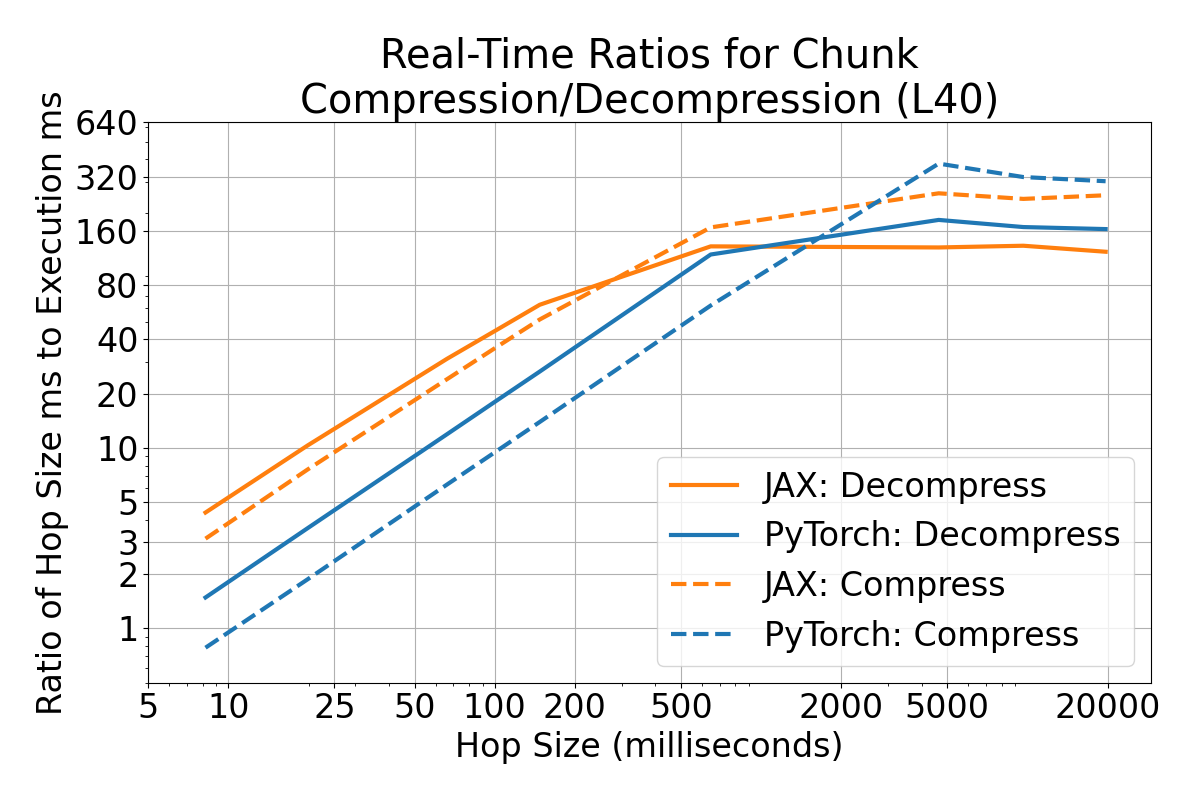}
  \end{minipage}
  \vspace{-0.5\baselineskip}
  \caption{Log-log plot of the unitless ratio of hop size to execution time.
  }
  \vspace{-1.5\baselineskip}
  \label{fig:real-time-ratio}
\end{figure*}

\subsection{Performance on RTX 2080}

For the RTX 2080 with 8 GB of memory, JAX outperforms PyTorch across the board. In this experiment, JAX is running in a Windows Subsystem for Linux (WSL) prompt with Python 3.10 versus a plain command prompt and Python 3.11 for PyTorch. We expect both of those conditions to give a slight advantage to PyTorch. For JAX, we use version 0.4.25 with Python 3.10 in a WSL command prompt. For PyTorch, we use version 2.2.0 with Python 3.11 in a Windows command prompt.

The decompression times have higher variance in JAX than in PyTorch. Since the mean decompression time for JAX is already lower, it would be interesting to find a way to reduce the variance and achieve an even lower mean.

\subsection{Performance on L40}

On a Lenovo ThinkSystem SR670 V2 with an Nvidia L40 GPU with 48 GB of memory, we use JAX 0.4.27, PyTorch 2.3.0, and Python 3.10. For a hop size smaller than 0.647 sec, JAX performs compression and decompression faster than PyTorch. For a hop size above 4.6 sec, PyTorch performs faster, and the crossover point lies between. For JAX, relative to PyTorch, it is interesting that the variance of the compression times is low, and yet the variance of the decompression times is high.

\subsection{Choosing the hop size}

If latency is not a concern, then the developer might want to pick a large window size compatible with their GPU memory capacity. For an RTX 2080, \figref{fig:real-time-ratio} shows that the ratio of hop size (in ms) to execution time (in ms) rises as a function of hop size. This is a measure of real-time performance. However, it ignores the execution cost of functions that are outside of the for-loops. For PyTorch, this would involve pyloudnorm, and for JAX it would involve jaxloudnorm. Similarly, if resampling were necessary, PyTorch would use Julius, and JAX would use our re-implementation of Julius.

\section{Discussion}\label{discussion}

\subsection{Faust-JAX}\label{faust-jax}

DAC-JAX creates research opportunities with the existing Faust-JAX pipeline. Faust is a domain-specific language for audio synthesis~\cite{orlarey2009faust}. Faust’s built-in libraries include functions for reverbs, compressors, oscillators, filters, ambisonics, and more.\footnote{\url{https://faustlibraries.grame.fr/}} Faust supports many compilation ``targets'' such as C/C++, Rust, WebAssembly, and recently{\textemdash}via a contribution by the author{\textemdash}Flax. A Google Colab in the DawDreamer module demonstrates how to write instruments and effects in Faust and transpile to Flax~\cite{braun2021dawdreamer}.\footnote{\url{https://github.com/DBraun/DawDreamer/blob/main/examples/Faust_to_JAX/Faust_to_JAX.ipynb}} It shows how to parallel-render synthesizer one-shots with many settings at different notes and velocities. It also uses gradient descent to optimize the automation of a low-pass filter's cutoff parameter.

For future research, we suggest training a language model on DAC-JAX tokens to create an embedding space of a corpus of music, and then using Faust with JAX to explore instrument and effects.

\subsection{Penzai}

With DAC-JAX, future research can now use the new Penzai JAX module to inspect intermediate values of audio neural networks.\footnote{\url{https://github.com/google-deepmind/penzai}} If DAC-JAX were used as the codec in a model like MusicLM or MusicGen (which would also need to be implemented in JAX), then Penzai could be used to inspect the attention heads of the transformer layers.

\section{Conclusion}

We present an open-source JAX implementation of the Descript Audio Codec (DAC). We match most important features of the original DAC including reusing weights, training with device parallelism, JIT-optimized inference, and command-line interfaces. On consumer hardware, DAC-JAX performs chunked compression and decompression faster than the original DAC. However, for the same experiment on a remote Linux cluster, JAX only performs faster on hop sizes smaller than 647 milliseconds. Further (more methodical) benchmarks should use different hardware (e.g., GPUs, TPUs, CPUs) and also measure the training speed.
\newpage

\bibliography{ISMIRtemplate}

\end{document}